\documentclass[preprint]{aastex}

\received{}
\revised{}
\accepted{}
\ccc{}
\cpright{}{}

\slugcomment{AJ in press, August 2004}
\shorttitle{X-ray Orion FF}
\shortauthors{}

\newcommand{\subsun}{\mbox{$_{\odot}$}}
\newcommand{\etal}{{\it et al.\/}}

\newcommand{\lx}{${\rm L_x}$}
\newcommand{\lbol}{${\rm L_{bol}}$}

\begin{document}

\title{Chandra X-ray observations of Young Clusters 
II. Orion Flanking Fields Data.
}

\author{ 
Solange V. Ram\'{\i}rez \altaffilmark{1},
Luisa Rebull \altaffilmark{1},
John Stauffer \altaffilmark{1},
Stephen Strom \altaffilmark{2},
Lynne Hillenbrand \altaffilmark{3},
Thomas Hearty \altaffilmark{4},
Eugene L. Kopan \altaffilmark{5},
Steven Pravdo \altaffilmark{4},
Russell Makidon \altaffilmark{6}, \&
Burton Jones \altaffilmark{7}.
}

\altaffiltext{1}{Spitzer Science Center, Mail Stop 220-06,
California Institute of Technology.}
\altaffiltext{2}{NOAO, Kitt Peak National Observatory}
\altaffiltext{3}{Astronomy Department, California Institute of Technology.}
\altaffiltext{4}{Jet Propulsion Laboratory}
\altaffiltext{5}{IPAC, California Institute of Technology.}
\altaffiltext{6}{Space Telescope Science Institute}
\altaffiltext{7}{Astronomy Department, California Institute of Technology.}

\begin{abstract}
We present results of Chandra observations of two flanking fields (FF) 
in Orion, outside the Orion Nebula Cluster (ONC).
The observations were taken with the ACIS-I camera 
with an exposure time of about 48 ks each field.
We present a catalog of 417 sources, which includes X-ray luminosity,
optical and infrared photometry and X-ray variability information. 
We have found 91 variable sources, 33 of which have a flare-like 
light curve, and 11 of which have a pattern of a steady increase or 
decrease over a 10 hour period.
The optical and infrared photometry for the stars identified as X-ray 
sources are consistent with most of these objects being pre-main 
sequence stars with ages younger than 10 Myr.
We present evidence for an age difference among the X-ray
selected samples of NGC 2264, Orion FF, and ONC, with
NGC 2264 being the oldest, and ONC being the youngest.

\end{abstract}


\section{INTRODUCTION}

The Orion Nebula is one of the best studied star formation regions 
in the Galaxy.
The Orion Nebula Cluster (ONC) designates the inner 3 pc ($\sim$ 20 arcmin)
from $\theta^{1}$C Orionis, and contains more than three thousand stars.
The core of the ONC, also known as the Trapezium Cluster \citep{tru31,her86},
comprises the stars located within 0.3 pc ($\sim$ 2 arcmin) of
$\theta^{1}$C Orionis, it has an inferred stellar density
of about 8000 stars pc$^{-3}$ and it is one of the
densest known region of star formation \citep{mcc94}.
Most of the ONC members are visible, primarily because the massive stars 
in the Trapezium Cluster have swept the gas from the molecular cloud 
creating a cavity facing towards us. 
The structure, dynamics and stellar content of the ONC have been extensively 
studied \citep{hil97,hil98,hil00,car00,luh00,ode01}. 
The stars in the ONC are less than 1 Myr old and their masses range 
from $<$0.05 M\subsun~ to $\sim$50 M\subsun ~\citep{hil97,hil00}. 
Studies of the outer regions of the Orion Nebula (Flanking Fields) have 
revealed the presence of young stars of ages between 1 and 3 Myrs old 
\citep{reb00,gul98,har98}, with an accretion disk fraction of about 40\% 
\citep{reb00}.
The Orion Nebula provides an excellent laboratory for studying pre-main 
sequence (PMS) stars, given its proximity (470 pc), its age and its stellar
diversity.

PMS stars are strong X-ray emitters; typical PMS stars have X-ray 
luminosities $10^1$ to $10^4$ above those observed in older main
sequence stars \citep{fei99}. 
The X-ray emission in low mass stars comes from magnetic reconnection. 
In the dynamo model, the strength of the magnetic field, and hence the 
X-ray activity, depends on the rate of differential rotation and on the 
depth of the outer convective envelope \citep{gil83,ros85}.
There is a clear relationship between rotation rate (period) and
X-ray luminosity (\lx) found in late-type stars in clusters as old as
NGC 2547 \citep[15-40 Myrs,][]{jef00} through the Hyades
\citep[$\sim$500 Myrs,][]{sta97}.
The ratio between the X-ray and bolometric luminosity, \lx/\lbol,
increases with increasing rotation rate, until the most rapidly
rotating stars reach a maximum X-ray luminosity (or saturation level)
such that \lx/\lbol $\sim 10^{-3}$
\citep[see ][and references within]{piz03}.
It is much less clear that rotation and \lx/\lbol~ are related in
younger clusters.
\citet{gag94} studied a 4.5 square degree region centered in the 
Orion Nebula, using X-ray data from the Einstein Observatory.
They found that at least 100 sources were associated with late-type
PMS stars, and their X-ray activity was not correlated with published
rotational periods and spectroscopic rotational velocities. 
\citet{gag95} used ROSAT X-ray data in a similar study in the Orion
Nebula, finding no dependence of X-ray activity on rotation.
More recently, the Chandra Observatory has provided significantly 
improved spatial resolution and sensitivity for X-ray observations.
\citet{fei02} reported X-ray observations of the ONC
using the Chandra Observatory ACIS-I detector. 
They found more than a thousand X-ray sources, 91\% of them associated 
with known stellar members of the cluster. 
\citet{fei03} also see no obvious correlation between rotation and \lx/\lbol~ 
for their ONC sample, and conclude that the X-ray generation mechanism
for young PMS stars must be different from that responsible in
young main sequence stars.   
\citet{fla03b} analyzed data for a number of young associations and 
clusters (including Orion), and agree that there is little correlation 
between \lx/\lbol~  and rotation at very young ages, but conclude 
that the data are consistent with a single physical mechanism, where 
the Orion-age stars are simply all at or near the saturation level.
We intend to use our Orion flanking field data and other new Chandra data
for NGC 2264 \citet{ram04} to help determine at what age the 
relationship between rotation and \lx/\lbol~ emerges.

In the present paper, we present results of Chandra observations
of two of the flanking fields (FF) in the Orion Nebula 
\citep[fields \#2 \& \#4 from ][]{reb00}, just outside the ONC. 
We discuss source detection, variability and \lx~ determination,
providing a catalog of 417 X-ray sources.
We present evidence that the X-ray selected sample of PMS
stars in the flanking fields in Orion are older than a similarly
selected sample in the ONC.
In paper III \citep{reb04}, we will discuss in more detail the 
relationships found here between rotation rate, mass accretion rate, 
disk indicators, and X-ray luminosity, in our field of NGC 2264 and
the two Orion flanking fields.  

\section{OBSERVATIONS}

Two flanking fields in Orion were observed with the Advanced CCD 
Imaging Spectrometer (ACIS) detector on board the 
{\it Chandra X-ray Observatory} \citep{wei02}.
The imaging array (ACIS-I) consists of four 1024$\times$1024 
front-side illuminated CCDs and covers an area on the sky of about 
17'$\times$17'.
The North Orion Flanking Field (NOFF) is centered at $5^h35^m19^s, 
-4\arcdeg48\arcmin15\arcsec$, about 36 $\arcmin$ ($\sim$ 5 pc, at 
a distance of 470 pc) north
of the Trapezium Cluster, and it was observed on 2002 August 26 
with a total exposure time of 48.8 ks.
The South Orion Flanking Field (SOFF), centered at $5^h35^m6^s, 
-5\arcdeg40\arcmin48\arcsec$, about 17 $\arcmin$ ($\sim$ 2.5 pc, at 
a distance of 470 pc)
south of the Trapezium Cluster, was observed on 2002 September 6 
with a total exposure time of 47.9 ks.
Figure ~\ref{dss} shows a 1$\arcdeg \times 1.5\arcdeg$ image of Orion from the
Palomar Digital Sky Survey \citep{rei91}, with both Chandra ACIS-I 
fields of view marked as boxes.
The fields were selected to maximize the number of stars
in the field of view with known periods from \citet{reb01} and 
with minimal overlap with other contemporaneous Chandra Orion 
observation \citep{fei02,fla03b}.

\subsection{Data Preparation}\label{data}

Data analysis was performed in the same manner as described in \citet{ram04},
where we discussed similar observations of a field in NGC2264.
In summary, we started with the Level 1 processed event list,
applied charge transfer inefficiency (CTI) correction as described in
\citet{tow00}, then filtered by ASCA grades; by time intervals; 
and by background flaring due to solar activity. 
Removal of the background flares identified by the latter step reduced 
the exposure time to 46.9 ks for the NOFF and to 46.3 ks for the SOFF.
Finally, the energy range was restricted to 0.3 -- 10 keV.
The filtering process was done using the Chandra Interactive Analysis 
of Observations (CIAO) 
package\footnote{http://cxc.harvard.edu/ciao/index.html}.
Figures ~\ref{orion_N}  and  ~\ref{orion_S} show the ACIS-I image of 
the filtered observations for the NOFF and the SOFF, respectively.

\subsection{Source detection}

X-ray sources were identified using the {\it wavdetect} tool within the
CIAO package. 
The {\it wavdetect} tool works well in detecting X-ray sources in
crowded fields.
This tool performs a Mexican hat wavelet decomposition and
reconstruction of the image as described in \citet{fre02}. 
We used wavelet scales ranging from 1 to 16 pixels in steps of $\sqrt{2}$,
and the default source significance threshold of $1\times10^{-6}$.
The {\it wavdetect} tool was run separately in the four ACIS-I CCDs images
for both fields and it produced an original list of 256 sources for the
NOFF and 280 sources for the SOFF (see Sec~\ref{xray_phot}). 

\subsection{Astrometric Alignment}

The positions of the sources obtained by {\it wavdetect} were correlated
with 2MASS positions.
Each X-ray source was manually checked to confirm that the 2MASS 
counterpart lay within the radius for count extraction (as defined in
\citet{ram04}).
We checked the astrometry of the Chandra observations using all X-ray 
sources located within an off-axis angle ($\phi$) less than 5 arcmin 
that have 2MASS counterparts.
A total of 85 NOFF sources and 70 SOFF sources meet this criteria.
We determined a mean offset in R.A. of $-0.09\arcsec \pm$ 0.02 and a mean 
offset in DEC of $-0.14\arcsec \pm$ 0.03 between the NOFF Chandra and 2MASS 
coordinates and a mean offset in R.A. of $0.09\arcsec \pm$ 0.03 and a mean
offset in DEC of $-0.28\arcsec \pm$ 0.03 between the SOFF Chandra and 2MASS
coordinates.
The Chandra positions were corrected by these mean offsets 
to put the X-ray sources in the same reference frame as their 2MASS
counterparts.

\section{RESULTS}

\subsection{X-ray Photometry}\label{xray_phot}

X-Ray aperture photometry was initially performed on the 536 sources 
detected by the CIAO tool {\it wavdetect}. 
The radius for count extraction, $R_{ext}$, and the annulus for background
determination were defined in the same manner as in \citet{ram04}.
The X-ray counts ($C_{extr}$) and background counts were extracted 
using the CIAO tool {\it dmextract}.
We computed the background in counts$\times$arcsec$^{-2}$ as a function
of off-axis angle ($\phi$), performing a 3$\sigma$ rejection fit to
avoid background counts from annuli that have other sources within.
The computed background was constant as a function of $\phi$, and 
it had a value of $B$=(0.072$\pm$0.024)counts$\times$arcsec$^{-2}$
for the NOFF and a value of $B$=(0.059$\pm$0.021)counts$\times$arcsec$^{-2}$
for the SOFF. 
The net count, {\it N.~C.} and the count rate, {\it C.~R.} were computed 
for each source as in \citet{ram04}.

We carefully inspected the light curves of all the sources (see 
Sec.~\ref{variability}) and their appearance in the image of the 
Chandra field of view. 
A total of 119 (22\%) sources were rejected from the original list: 
107 sources had light curves consistent with cosmic ray afterglows,
5 sources were detected twice, since {\it wavdetect} was run separately 
in each CCD, and 7 sources were located at the edge of the field of view.
Of the 107 light curves having a cosmic ray shape, 106 (99\%) contain
cosmic ray afterglows flagged by the pipeline.
Our final list of 417 X-ray sources (203 from the NOFF, 214 from the SOFF) is 
provided in Table~\ref{tab_xray}.

\subsection{X-ray Luminosities}\label{xray_lum}

We selected all sources with more than 500 net counts, extracted
their spectra and fit them to measure their X-ray fluxes. 
A total of 24 sources from the NOFF and 20 sources from the SOFF meet 
this criteria and they are listed in Table~\ref{tab_spectra}.
The spectra were extracted within $R_{ext}$ using the CIAO tool 
{\it dmextract}. 
The spectra of the NOFF and the SOFF sources are shown in
Figures~\ref{spectra_N} and ~\ref{spectra_S}, respectively.
We used the same spectral fitting process, as described by \citet{ram04},
with the Redistribution Matrix Files (RMF) provided by the CTI corrector 
\citep{tow00}, Auxiliary Response Files (ARF) created by the CIAO tool 
{\it mkarf} and later corrected for the ACIS low energy quantum 
efficiency degradation.
We adopted a photoelectric absorption model ($xswabs$), which uses 
Wisconsin cross sections from \citet{mor83}.
This model has one parameter that is the equivalent hydrogen column 
density ($nH$). 
The hydrogen column density was fixed to a value of 0.08$\times 10^{22}
{\rm cm^{-2}}$ to match an extinction value of $A_V$ = 0.41, 
which is the most likely value of the observed extinction towards
both Orion flanking fields \citep[see ][for more discussion]{reb00}.
As discussed in \citet{ram04}, the error in \lx~ that comes from 
fixing $nH$ is negligible.
Finally, we adopted a thermal emission model ($xsmekal$)
based on the model calculations of \citet{mew85,mew86}, and \citet{kaa92} 
with Fe L calculations by \citet{lie95}.
This model includes line emissions from several elements.
The remaining two parameters in the model are the plasma temperature 
(kT) and a normalization factor. 
As in the analysis of NGC 2264 Chandra observations \citep{ram04},
we used a two temperature model to fix the X-ray spectra.
The spectral parameters obtained from the two plasma temperature models
are listed in Table ~\ref{tab_spectra}.
The X-ray flux for the brightest sources is determined from 
the best spectral model derived from the mean model parameters. 
We computed mean plasma temperatures of (0.63$\pm$0.05) keV and
(2.6$\pm$0.3) keV for the NOFF sources and mean plasma
temperatures of (0.77$\pm$0.05) keV and (2.9$\pm$0.2) keV for the SOFF
sources. 
The mean plasma temperatures are held constant and the integration of
the best fit between 0.3 and 8.0 keV provides the X-ray flux.
The resulting X-ray fluxes are listed in Table ~\ref{tab_spectra}.

We use the X-ray fluxes of the bright sources to compute a X-ray 
flux weighted conversion factor between count rate and X-ray flux. 
We obtained a conversion factor of (6.58$\pm$0.13)$\times10^{-15}$
(erg/cm$^{2}$/s)/(counts/ks) for the NOFF sources and a conversion factor of
(6.72$\pm$0.20)$\times10^{-15}$ (erg/cm$^{2}$/s)/(counts/ks) for the 
SOFF sources.  
The values of these conversion factors are in reasonably good agreement
with the derived conversion factor for our similar X-ray
observation in NGC 2264
\citep[6.16$\pm$0.13$\times10^{-15}$ (erg/cm$^{2}$/s)/(counts/ks)][]{ram04}
and other published values \citep[see ][]{ram04}.
The X-ray flux for our catalog of X-ray sources in both Orion fields 
is listed in column 10 of Table~\ref{tab_xray}.
The X-ray luminosity,~\lx, listed in column 11 of Table~\ref{tab_xray}, 
is computed assuming a distance to Orion of 470 pc.

The limiting luminosity in our X-ray observations varies within the 
field of view, because of the variation of the PSF across the field. 
The faintest source is located at $\phi \sim 6 \arcmin$ and
it has a count rate of 0.08 counts/ks, corresponding to a X-ray 
luminosity of 28.15 at the distance of Orion. 
About 85 \% of our sources are located within $\phi = 8\arcmin$.
At that off-axis angle the limiting count rate has increased to
0.20 counts/ks, corresponding to a X-ray luminosity of log(\lx)=28.5 
at the distance of Orion. 
At $\phi = 10\arcmin$, we cannot detected X-ray sources fainter than 
0.45 counts/ks (log(\lx) = 28.9).
Therefore, we adopt a value of log(\lx)=28.5 dex as the limiting 
X-ray luminosity for our observations, keeping in mind that this
value holds for sources located within $\phi \sim 8 \arcmin$.

  
\subsection{Variability}\label{variability}
 
\begin{figure}
\epsscale{0.7}
\plotone{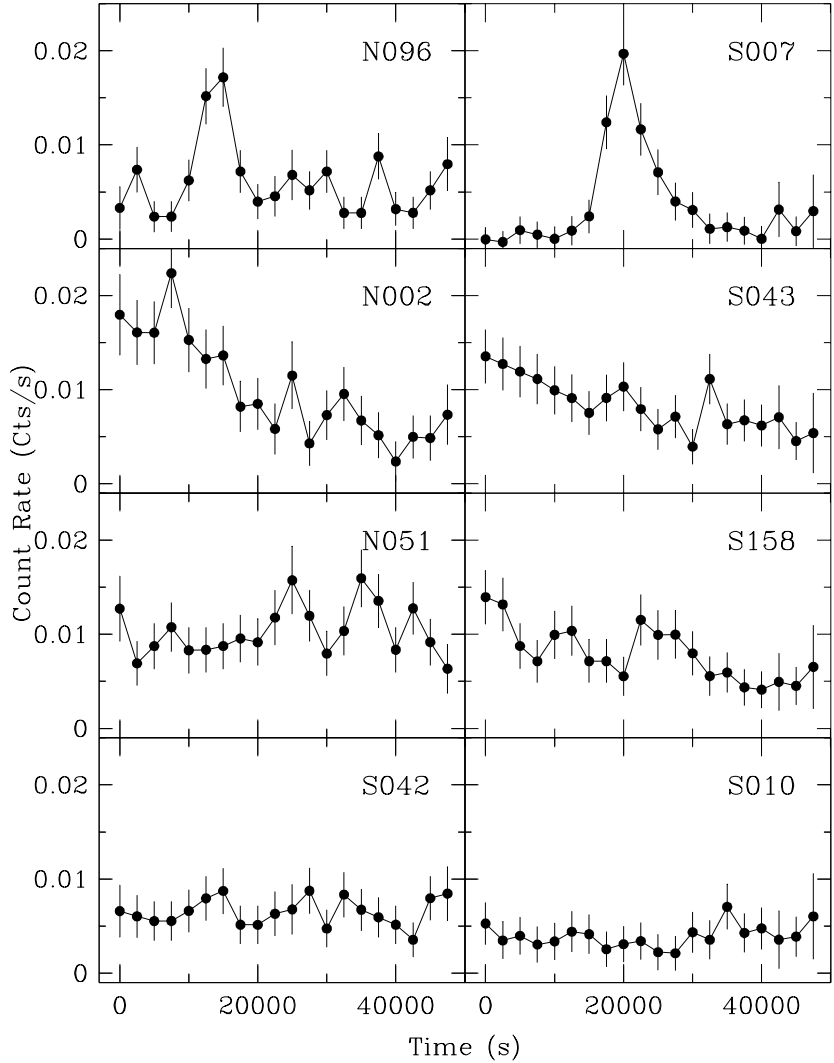}
\caption[light_curves.ps]{Examples of light curves for eight X-ray sources
of comparable luminosity from our sample. 
The two sources in the top panels show flare-like light curves. 
The two sources on the middle-top panels show a steady decrease in
their light curves.
The two sources on the middle-low panels are variable, according to our
statistical test, and 
the two sources in the bottom panels are constant.
The sources in the left panels are from the NOFF and
the sources in the right panels are from the SOFF.
There are 91 variable sources in our Orion sample of 417 X-ray objects; 
33 of them show a flare-like light curve. 
\label{light_curves}}
\end{figure}

Light curves were determined for all 417 sources detected by {\it wavdetect} 
using the CIAO tool {\it lightcurves} with a bin time of 2500 s.
The statistics of the light curves of the sources of our X-ray catalog
were obtained using the Xronos
\footnote{http://heasarc.gsfc.nasa.gov/docs/xanadu/xronos/xronos.html} 
tool {\it lcstats}. 
This provides, among other values, the probability of the light curve 
being constant, $P_{c}(\chi^{2})$, as derived from the Chi-square value.
The $P_{c}(\chi^{2})$ values are listed in column 12 of Table~\ref{tab_xray}.
The light curves of the sources with $P_{c}(\chi^{2}) < $ 90\% were 
analyzed further, since they are the most likely to be variable.
We defined a variable source as those having 2500 s bin light curves 
with $P_{c}(\chi^{2}) < $ 90\%, and 5000 s and 7500 s bin light curves
with reduced $\chi^{2} > $ 2.5, as in \citet{ram04}.
There are 91 variable sources that meet this criteria.
Variable sources are marked with a 'v' in column 13 of 
Table~\ref{tab_xray}.
There are 33 variable sources which show a flare shape, defined as a 
rapid increase and a slow decrease in the X-ray flux.  
Variable sources with a flare-like light curve are
marked with an additional 'f' in column 13 of Table ~\ref{tab_xray}. 
There are sixteen sources that show a possible flare pattern, described
as an increase in X-ray flux happening at the end of our observations
or a decrease in X-ray flux occurring at the beginning of our observations.
Variable sources with a possible flare pattern are marked
with an additional 'p' in column 13 of Table~\ref{tab_xray}.
We have detected a fraction of about 11\% flaring sources (8\% excluding
possible flares) in both Orion flanking fields. A comparable
fraction (about 8\%) was observed in a similar length observation in NGC 2264
\citep{ram04}.
There are 11 sources that show a steady increase or decrease in 
their X-ray flux.
Those sources are marked with an additional 's' in column 13 of
Table~\ref{tab_xray}. A similar pattern is observed in two NGC 2264
sources \citep{ram04} and three X-ray IC 348 sources \citep{pre02}.
These patterns might be understood by rotational modulation of X-ray flares
\citep{ste99}.
In Figure ~\ref{light_curves}, we show a selection of light curves
of sources of comparable luminosity.
In the top panels, we show two light curves with a flare shape;
in the middle-top panels, there are two sources that show a steady 
decrease in their X-ray flux;
in the middle-bottom panels, there are two variable light curves.
Finally in the bottom panels, two constant light curves are plotted.

\section{DISCUSSION}

\subsection{Description of the optical/infrared catalog}\label{catalog}
 
We correlated the X-ray sources with a catalog of optical and near-IR
stars in the Orion region; see Tables~\ref{tab_optical} and ~\ref{tab_xids}.  
We constructed the catalog by merging $\sim$30 published catalogs of 
optical and infrared photometry, spectral types, previous X-ray detections, 
proper motion surveys, periods, and spectroscopic projected rotational 
velocities ($v \sin i$).  
Although our X-ray data presented are located in two flanking fields 
\citep[\#2 and \#4 from][]{reb00}, we constructed a catalog over this whole 
region in an attempt 
to increase the number of stars with known \lx~ and optical counterparts. 

The optical photometry were taken from \citet{reb00}, \citet{hil97}, 
\citet{sta99}, \citet{wol03}, and \citet{gag95}, in that order.
For near-IR photometry, \citet{car01} averaged over their light curves, 
presenting the most reliable average $JHK$ magnitude.  
Magnitudes for the remaining stars then came from 2MASS,
\citet{mue02}, \citet{hil98}, and  \citet{hil00}, in that order.
Spectral types were taken in this order: 
\citet{reb00}, \citet{hil97}, \citet{her88}, \citet{dun93},
\citet{edw93}, \citet{smi83}, \citet{wal69,wal83}, \citet{gag94},
\citet{gag95}, and \citet{wol03}.
\citet{gag94} and \citet{gag95} collected much of the optical and 
spectroscopic literature for stars in this region, reporting on X-ray 
detections using Einstein and ROSAT, respectively.  
\citet{str90} analyzed Einstein data for some stars in this region.  
The largest X-ray data base, and the one obviously most similar to our 
own data, is \citet{fei02}; these Chandra ACIS X-ray detections are in the
ONC region; their fields have marginal spatial overlap with the observations
reported here.
Finally, there are several surveys in this region devoted specifically to
obtaining rotation data.  
Periods were taken from \citet{reb01}, \citet{sta99}, \citet{her00,her01}, 
\citet{car01}, \citet{gag95}, in that order.
Spectroscopic rotational velocities came from \citet{rho01}, \citet{wol03}, 
\citet{dun93}, \citet{smi83}, \citet{gag95}, \citet{har86}, and finally
\citet{wal80}.

\subsection{Comparison of optical/infrared and X-ray Catalogs}\label{ccatalog}

We found that 356 (85\%) of our 417 X-ray sources have optical and/or
infrared counterparts (130 (31\%) have known periods).  
Each X-ray source was manually checked to confirm that the optical and/or
infrared counterparts lay within the radius for count extraction ($R_{ext}$,
as defined in \citet{ram04}).
The 356 sources and their corresponding optical and/or infrared
photometry are listed in Table~\ref{tab_optical}.
Other names of the sources given by the different catalogs used in our
compilation are listed in Table~\ref{tab_xids}.

\begin{figure}
\epsscale{0.7}
\plotone{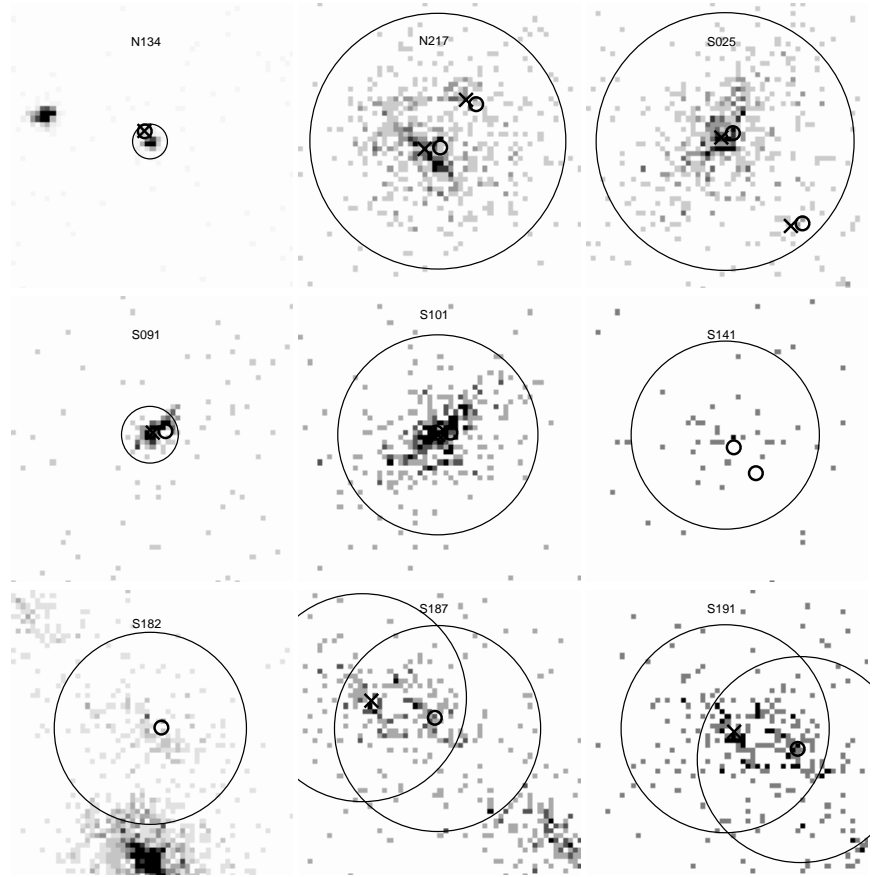}
\caption[find_chart.ps]{Finding charts for 9 sources that show 
several optical ($circle$) and/or infrared ($cross$) counterparts 
within their extraction circle ($thin~circle$) or have contamination 
from nearby X-ray sources.
\label{find_chart}}
\end{figure}

There were cases in which two optical/near-IR counterparts were located
within $R_{ext}$, which is a measure of the size of the PSF in the
Chandra field.
This was the case for 9 X-ray sources, shown in Figure~\ref{find_chart}.
These sources are marked with a 'c' in Table~\ref{tab_xray}
and are matched to the closest counterpart in each case.
Source N134 seems to be have two components separated by $\sim$ 1''.
The optical (Par 1970) and 2MASS counterparts coincide with the 
weaker component. Par 1970 is a B2 star, which should be weak in X-rays. 
Source N217 has two pairs of optical and 2MASS counterparts 
separated by $\sim$ 6.5''.
Source S025 has two pairs of optical/2MASS counterparts within
the extraction circle. 
One pair is at the edge of the extraction circle and the other 
within 1.3'' of the X-ray position. 
Source S091 is an extended source, possibly with two components.
Source S101 has two optical and one infrared source within 
the extraction circle. 
Source S141 has two optical sources within the extraction circle. 
Source S182 has a large aperture and its extraction circle contains
emission from a nearby X-ray source. 
Sources S187 and S191 are separated by 5.5'' at an off axis angle of
about 8.5'. Their circles of extraction ($R_{ext} \sim$ 10'')
overlap considerably. Source S187 has an optical counterpart (Par 1897),
while source S191 has an infrared counterpart (2MASS J05351696-0532464).
Optical and/or infrared spectroscopy of these objects may help determine
the true counterpart of these X-ray sources.

\begin{figure}
\epsscale{0.7}
\plotone{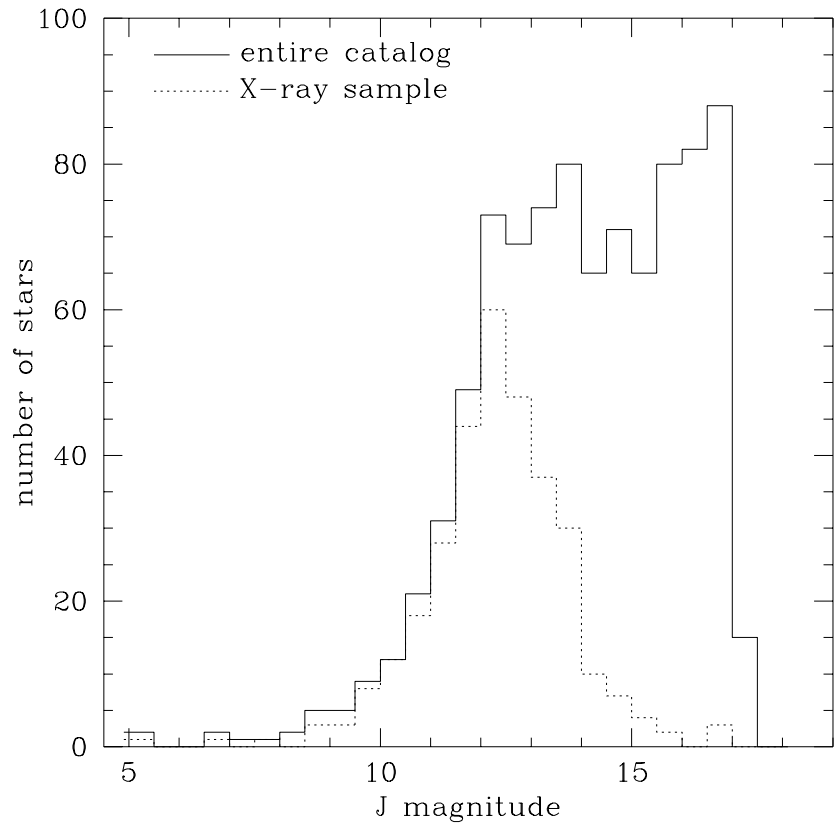}
\caption[j_histogram.ps]{$J$ magnitude histogram of all the sources
with $J$, $H$, and $K$ magnitudes in our Chandra fields.
The solid line corresponds to all the sources, and the dotted line
corresponds to all the sources with X-ray counterparts.
All the X-ray sources should have been matched to sources in our catalog
if they are associated with stars earlier than M4 at the distance and age 
of Orion.
\label{j_histogram}}
\end{figure}

There are 902 stars in our optical/infrared catalog with 
$J$, $H$, and $K$ photometry and with positions inside the field 
of view of our Chandra observations. 
Among those 902 stars, 316 (35\%) have X-ray Chandra counterparts.
Figure~\ref{j_histogram} shows a $J$ magnitude histogram of all 
the 902 stars with $J$, $H$, and $K$ photometry and positions 
inside the Chandra field (solid line) and the histogram 
of the 316 X-ray Chandra counterparts (dotted line).
The completeness limit of the infrared sample is that of 2MASS.
We can see in Figure~\ref{j_histogram}
that our infrared sample goes deeper than the sample of stars
with X-ray counterparts. 
A similar behavior is seen in the optical sample histogram. 
This means that all the X-ray sources should have been matched to sources 
in our optical/infrared catalog if they are associated with stars
earlier than M4 at the distance of Orion in the absence of extinction.

Among the Orion X-ray sources in our catalog, 130 (31\% of the 417 X-ray source)
have known periods from \citet{reb01}.
Among the Orion X-ray sources with known periods, 46 (35\% of the 130
X-ray sources with known periods) are classified as variable X-ray sources,
and 24 (18\% of the 130 X-ray sources with known periods) show a flare or
possible flare pattern.
Nearly 52\% of the 46 flaring X-ray sources are periodic in
optical wavelengths, while only 31 \% of all the X-ray sources show a
periodic optical variation.
There are twelve sources with short enough optical periods so that
a significant fraction of the period is observed in the time length of
the X-ray Chandra observations. However, the low signal to noise of the
light curves and the presence of X-ray flares prevent us from drawing any
conclusions regarding the presence or absence of rotational modulation
of the X-ray flux.

There are 61 X-ray sources that do not have optical or infrared 
counterparts and they are marked with a 'x' in column 13 of 
Table~\ref{tab_xray}. 
Based on the limiting magnitude of our optical catalog
($V\sim$ 20 mag), the X-ray to optical flux ratio ,$f_X/f_V$,
of the sources with only X-ray detection is $\ga 10^{-2.3}$. 
Therefore, these sources could be either late M dwarfs or 
extragalactic objects \citep{sto91}.
Given the presence of a dark cloud behind Orion,
the detection of background M dwarfs in X-rays is unlikely,
so, if these sources detected only in X-rays are
M dwarfs, then they should be $r \leq$470 pc.
We have identified 17 of these sources detected only in X-rays
with sources in a deep 2MASS image (Kopan 2004,
priv comm, $\sim$ 2 magnitudes
deeper than the 2MASS All Sky Point Source Catalog)
reducing the number of sources detected only in X-rays to 44.
Based on the limiting flux of our X-ray sample, 
the maximun X-ray luminosity of M dwarfs \citep{fle95},
and the limiting magnitude of the deep 2MASS image, 
we determined that all foreground M dwarfs present in our X-ray 
sample should have been detected in the deep 2MASS image.
Source counts in the Chandra South and North Deep Fields
\citep{ros02,bra01} predict the presence of about 20 AGN in 
each ACIS field of view at the flux limit of our observations 
and assuming a Galactic extinction of $A_V \sim$ 11 magnitudes
(from NED extinction 
calculator\footnote{http://nedwww.ipac.caltech.edu/forms/calculator.html}).
\citet{fei02} found 101 X-ray sources with no optical counterparts
in their 82 ks Chandra observation of the ONC, all highly concentrated 
towards the center of the ONC. 
They argued that most of those sources were indeed heavily embedded 
protostars or young stars seen behind the cloud.
They estimated that the extragalactic contamination cannot be more 
than 15 objects, due to the high extinction of the ONC behind the stars. 
In our case, the extinction in front of the Orion flanking field 
stars is significantly
lower than on the ONC \citep{reb00} and our only X-ray sources seem to be 
concentrated in areas of low extinction determined from a CO map of the
area \citep{bal87} and $H-K$ color of 2MASS sources in those areas.
Thus, we believe that the most likely explanation is that most 
of the objects detected only in X-rays are active galaxies.
Deeper infrared photometry or deeper X-ray observations may help 
determine the true nature of these sources.

There are 17 stars for which rotation and spectral type
information exist in our optical/infrared catalog, and
for which no X-ray counterpart was found.
In order to allow us to use these stars in the next paper,
we determined upper limits for their X-ray luminosity in the
same manner as \citet{ram04}.
The upper limit for the X-ray luminosity was computed
assuming a distance to Orion of 470 pc.
The upper limit results are listed in Table~\ref{tab_upp_lim}.

\subsection{Color-Color diagram.}

In Figure~\ref{color_color}, we have plotted the $J-H$, $H-K$
color-color diagram of all the infrared sources in the field
of view of our Chandra observation. 
Most of the X-ray sources with infrared colors are located near
the locus of the classical T-Tauri (CTT) stars \citep{mey97}. 
The remaining sources are located within the reddening vectors
of late main sequence stars and CTT stars.

\subsection{Color-Magnitude Diagrams.}

\begin{figure}
\epsscale{0.7}
\plotone{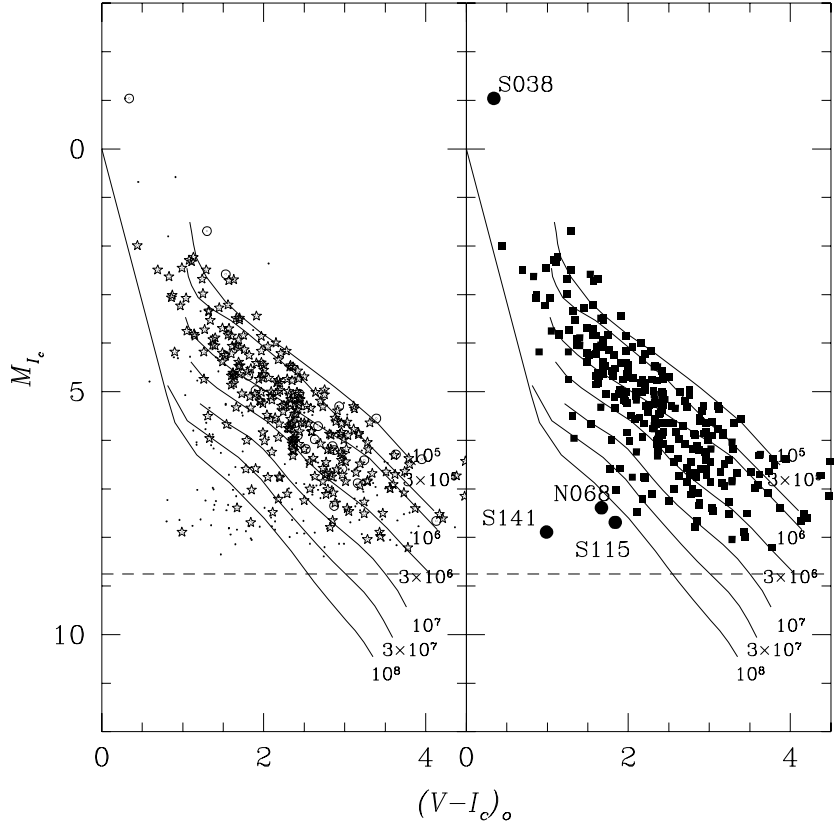}
\caption[cmd1.ps]{Dereddened optical color magnitude diagrams of 
sources located in the fields of the ACIS camera. 
In the left panel, $open~circles$ denote stars with X-ray counterparts 
and ${\rm L_x/L_{bol} < 10^{-4}}$, $stars$ denote sources with X-ray 
counterparts and ${\rm L_x/L_{bol} \geq 10^{-4}}$ , and the $dots$ mark 
the position of stars without X-ray counterparts.
In the right panel, only optical sources with X-ray counterparts are plotted.
The dashed line corresponds to $M_{I_{c}}$ = 8.75 mag., which is the lower 
limit for a low mass star rotating at the saturation level
(log(\lx/\lbol)=--3).
The solid lines denote isochrones from \citet{dan98}.
\label{cmd1}}
\end{figure}

There are 445 stars in our optical/infrared catalog with
$I_c$ and $V$ photometry and with positions inside both fields
of view of the Chandra observations.
Among those 445 stars, 280 (63 \%) have X-ray Chandra counterparts.
In Figure~\ref{cmd1}, we have plotted ($V-I_c$)--$M_{I_{c}}$
color magnitude diagrams of the optical sources in the fields 
of view of our Chandra observations. 
The dereddened $V-I_c$ color and the absolute magnitude $M_{I_{c}}$ were 
obtained assuming an average extinction of $A_V$ = 0.41 \citep{reb00}, and
dereddening relationships from \citet{fit99} and \citet{mat90}.
In both panels, we have also plotted isochrones from \citet{dan98},
as a reference.
The dashed line corresponds to $M_{I_{c}}$ = 8.75 mag., which
is the lower limit for a low mass Orion member rotating at the saturation 
level (log(\lx/\lbol)=--3) with log(\lx)=28.5 (limit luminosity of our X-ray sample). 
This means that the most slowly rotating stars earlier than M3 with 
optical counterparts should be detected by our X-ray sample.

Most of our X-ray sources are younger than $10^{7}$ years
at the distance of Orion, and furthermore the X-ray sources are
heavily concentrated between the $10^{5}$ and the $3 \times 10^{6}$ 
year old isochrones with respect to the general population. 
Stars that are strong X-ray emitters are likely
to be young stars and therefore true members of the Orion
cluster. In \citet{reb00}, as a result of the lack of
membership information for stars this far from the
Trapezium, a provisional membership criterion was defined.
A line was drawn through an empirically-discovered gap
nearly coincident with the 3 Myr isochrone to divide the
stars clumping above it in a locus (likely members) from the
ones below it (likely field stars).  This approximation was
supported in \citet{reb01} by the locations of the periodic
stars, also likely young and therefore members.  The
clustering of the X-ray sources above the 3 Myr isochrone is
additional confirmation that stars in the locus are more
likely to be true members of the cluster than stars below
this locus.

There is one star (S038) with $M_{I_{c}}$ brighter than 0 magnitude. 
Its optical counterpart, Par 1605, has a spectral type of A0V.
It has log(\lx)=30.1 and \lx/\lbol=$10^{-5.7}$. 
X-ray emission from late B stars and early A stars is rare, since
they are in the transition between wind driven X-ray emission from
hot stars and magnetically driven coronae from late type stars.
It is often speculated that the X-ray emission observed in late B and early A 
stars comes from unseen late type companions \citep{ste03,dan02},
that explanation could apply to Par 1605.

\begin{figure}
\epsscale{0.7}
\plotone{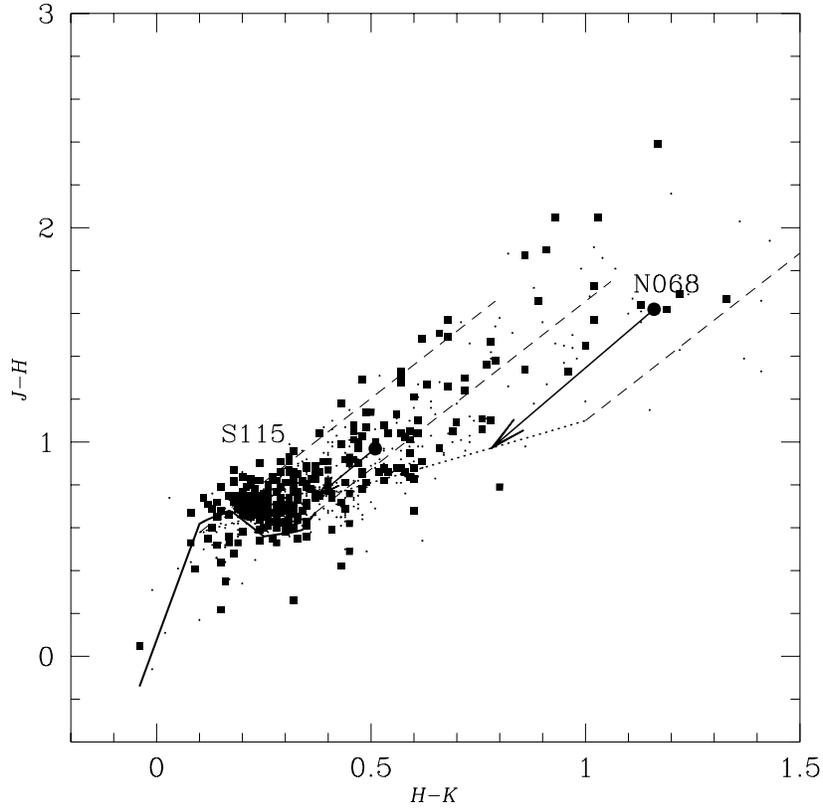}
\caption[color_color.ps]{Observed infrared color-color diagram of sources 
located in the fields of the ACIS camera. The $filled~squares$ denote stars 
with X-ray counterparts, and the $dots$ mark the position of stars 
without X-ray counterparts.
The $thick~line$ marks the location of main sequence colors, 
the $dotted~line$ the locus of CTTs from \citet{mey97},
and the $dashed~lines$ are reddening vectors; the length corresponds
to $A_V = 10$ magnitudes.
\label{color_color}}
\end{figure}

\begin{figure}
\epsscale{0.7}
\plotone{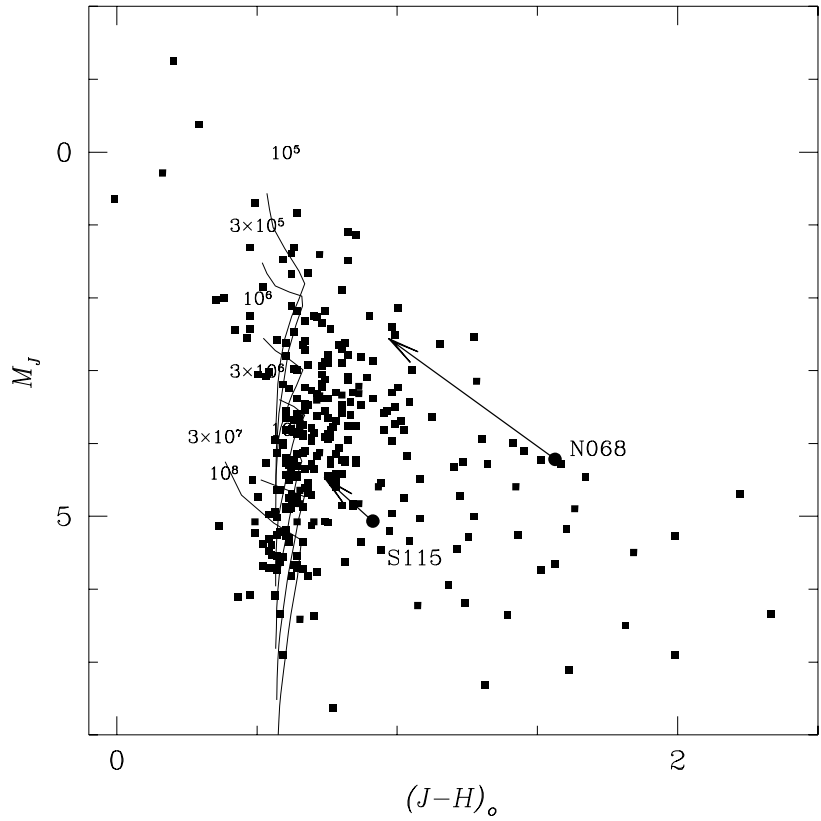}
\caption[cmd4.ps]{Dereddened infrared color magnitude diagram of sources located
in the fields of the ACIS camera. Only infrared sources with X-ray counterparts 
are plotted. 
The solid lines denote isochrones from \citet{dan98}.
\label{cmd4}}
\end{figure}

There are 3 X-ray sources that appear to be much older than $10^{7}$ years.
One of them (141S), has unreliable photometry because of background
contamination from a nearby star. The other sources (N068 and S115) 
lack both spectral type and membership information.
If we deredden sources N068 and S115 (see Figure~\ref{color_color})
along reddening vector towards
the locus of CTT stars, we derive a $A_V$ value of 5.9 magnitudes
for source N068 and $A_V$ value of 2.1 magnitudes for source S115.
The locations of sources N068 and S115 are indicated along with their
dereddened positions in Figure~\ref{color_color}.
In Figure~\ref{cmd4}, we have plotted the $(J-H)-M_J$ color magnitude
diagram of all the infrared sources with X-ray counterparts, with 
isochrones from \citet{dan98}. 
The locations of sources N068 and S115 are indicated
along with their possible dereddened position.
It is possible that these two objects are the youngest and most embedded of
our X-ray sources, if they are indeed members of the Orion complex. 
The unusual optical colors of these sources could be explained
by scattered light from an edge on disk.

\subsection{Age Difference.}

\begin{figure}
\epsscale{0.7}
\plotone{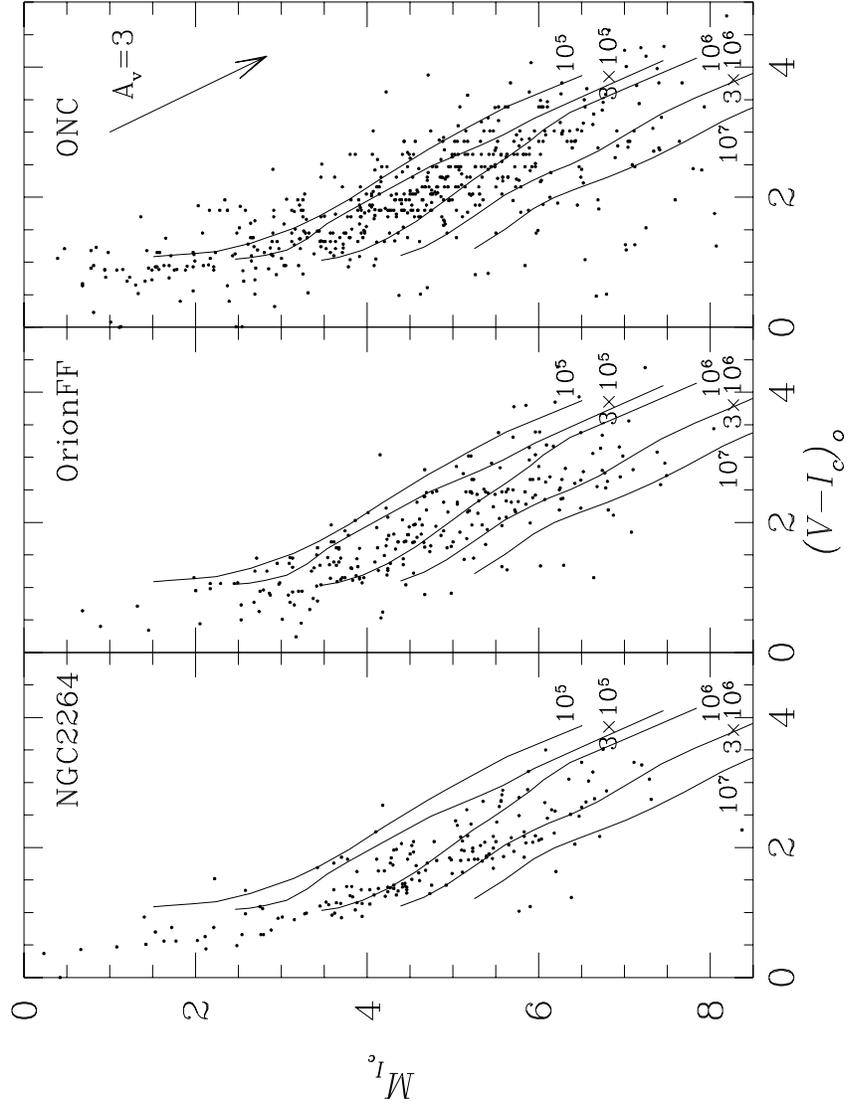}
\caption[xcmd.ps]{Dereddened optical color magnitude diagrams of
X-ray selected samples in NGC 2264 (left panel), Orion Flanking Fields
(central panel), and Orion Nebula Cluster (ONC, right panel).
The solid lines denote isochrones from \citet{dan98}.
The range in color selected for age comparison is 
1.0 $ < (V-I_c)_o < $ 2.5.
\label{xcmd}}
\end{figure}

Figure~\ref{xcmd} shows the dereddened color magnitude diagram
for three X-ray selected samples.
On the left panel, we plot the NGC 2264 sample from \citet{ram04},
which includes 201 stars showing X-ray emission.
The central panel shows Orion FF data from the present work.
Finally, the right panel shows 584 X-ray ONC sources \citep{fei02} that
have optical counterparts from our database. 
The stars from the three samples have been dereddened using the 
following criteria.
If the star has a known spectral type, we have used $A_I$ derived from 
the spectra. 
For the remaining stars, we have derived $A_I$ from $JHK$ photometry,
assuming that the stars are PMS stars. We use $A_I$ from infrared colors
if this is greater than one magnitude. 
If not, we use the median $A_I$ for each field ($A_I$=0.25 for NGC 2264,
$A_I$=0.25 for Orion FF, and $A_I$=1.31 for the ONC, Herbig \& Terndrup 1986). 
This criterion is reasonable for fields with variable reddening, as the
ONC, where $A_I$ can be as high as 4 magnitudes.

\begin{figure}
\epsscale{0.7}
\plotone{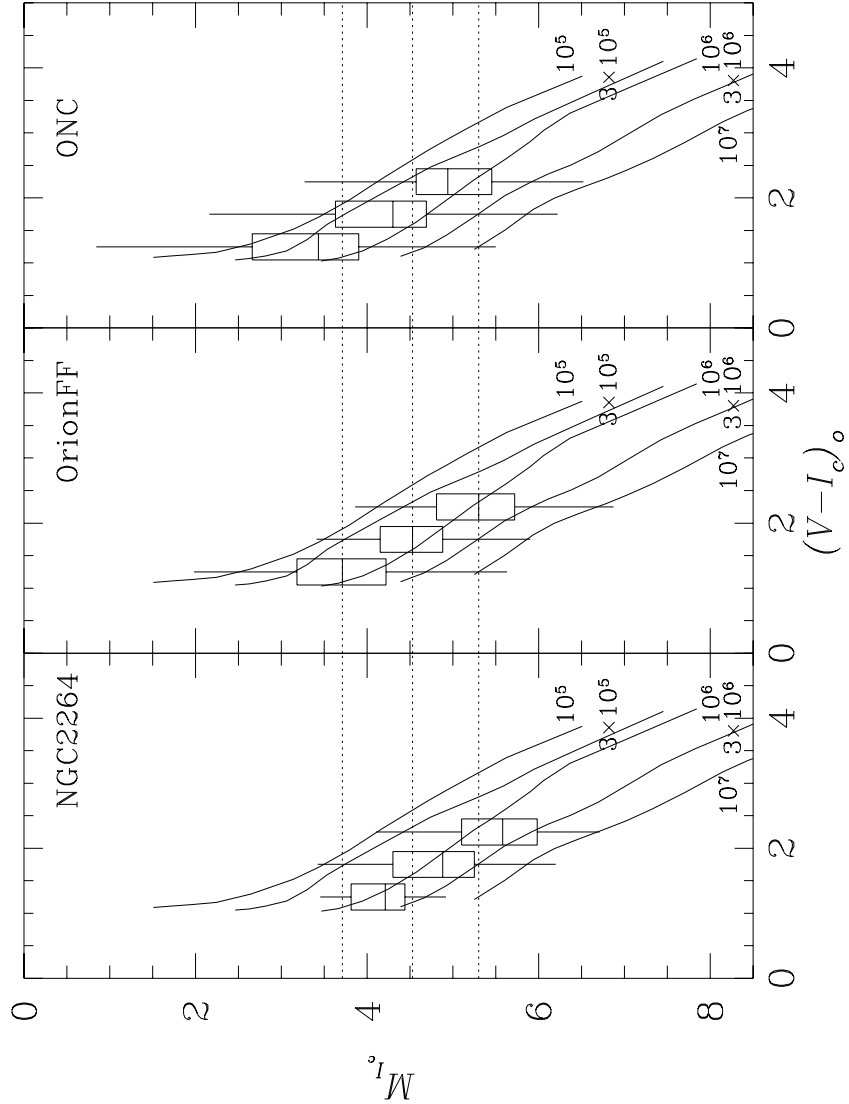}
\caption[xcmd_box.ps]{Dereddened optical color magnitude diagrams of
X-ray selected samples in NGC 2264 (left panel), Orion Flanking Fields
(central panel), and ONC (right panel), plotted as
statistical boxes. Each box corresponds to a range in $(V-I_c)_o$ of 0.5 
magnitudes. 
The central horizontal line of each box is the median $M_{I_c}$ in the
respective color range, the bottom and the top lines show its 
inter--quartile range and the vertical lines coming out of the box 
mark the position of the adjacent points of the sample.
The horizontal dotted lines show the median $M_{I_c}$ for the Orion
Flanking Field sample. 
There is a consistent age difference among the three X-ray samples.
The Orion Nebula Cluster sample is the youngest and the NGC 2264 
sample is the oldest.
\label{xcmd_box}}
\end{figure}

There are several notable differences among the shown CMDs. 
NGC 2264 stars appear to be more compactly distributed than the ONC stars. 
There are fewer high mass stars in the Orion FF sample than in the
ONC sample. 
The NGC 2264 sample appears to be the oldest of the three samples.
One way to quantify these differences is through the use of statistical
boxes. 
We have divided each X-ray selected sample into smaller samples
containing stars of a determined color range of 0.5 magnitudes of width. 
For each color range we have determined a box, as defined by 
\cite{tuk77} \citep[see also ][]{cle93}.
The central horizontal line in each box is the median $M_{I_c}$
for the range in $(V-I_c)_o$ color, while the bottom and the top of the box
shows its inter--quartile range (containing 50\% of the sample), and the 
vertical lines coming out of the box mark the position of the 
adjacent points of the sample (most extreme values in the sample 
that are not more than 1.5 times the inter--quartile range).
In Figure~\ref{xcmd_box} we show the box plots for the three
samples together with isochrones from \citet{dan98}. We have only 
plotted the boxes that overlap with all three samples, are within 
the range of isochrones ($(V-I_c)_o > 1.0$), and are not subject to 
biases from the optical limit of the catalog or other intrinsic 
properties \citep[$(V-I_c)_o < 2.5$; ][]{reb01}. 
We have also plotted in the three panels horizontal dotted lines 
that correspond to the median value of the Orion FF boxes. 
The median, upper and lower quartile $M_{I_c}$ values are listed 
in Table~\ref{tab_boxes} for each of the $(V-I_c)_o$ color ranges 
and for three the samples of X-ray sources.
All the NGC 2264 medians are systematically lower than the Orion FF
medians, with a mean difference of 0.3 magnitudes ($\sigma$=0.1).
All the ONC medians are systematically higher than the Orion FF medians,
with a mean difference of 0.3 magnitudes ($\sigma$=0.1). 
The effect of reddening is very small, since the reddening vectors go 
almost parallel to the isochrones. 
If we used the median reddening for each field, we find the same 
mean differences in the median $M_{I_c}$ for each color range.
If we used a $A_V$ one magnitude higher or one magnitude lower,
we get again the same differences in the median $M_{I_c}$.

The observed differences can be explained by a difference in age among
the three samples. 
Using the \citet{dan98} isochrones, we derive
an age of $\sim$2 Myrs, $\sim$1 Myrs, and $\sim$0.5 Myrs for the 
X-ray selected samples of NGC 2264, Orion FF, and ONC, respectively.
Alternatively, assuming vertical Hayashi tracks, stellar luminosities
scale as $L\sim t^{-2/3}$ \citep{har98}. Thus, differences in 
luminosity of 0.3 magnitudes corresponds to ratios in age of a factor
of $\sim$1.5. 
Searching for young stars in X-rays is known to be a efficient 
way of finding young stars \citep[for example, the RASS, e.g., ][]{neu95}.
Does the age difference for the {\bf{X-ray selected}\rm} stars in 
the three regions apply as well for all low mass members in the 
three regions?  
Our own belief is that X-ray selection does not bias YSO population
statistics in terms of age and mass \citep[e.g. ][]{fei02} and 
therefore our age estimates should be valid for the entire set of stars 
in the three regions.  
We caution however that X-ray selection may result in bias for other 
physical parameters, such as rotation or disk properties 
\citep[e.g. ][]{fla03a,fla03b}.
In order to provide some support to our assertion concerning X-ray selection 
bias and age, we have constructed plots like Figure~\ref{xcmd_box} for low 
mass members of the three clusters located within the Chandra field of view 
but not restricted to X-ray detection.   
We derived box-plots for the optical sample of likely members in
all three regions using two different reddening prescriptions -
one with a mean reddening for each field, and one where
we derived reddening exactly as for Figure~\ref{xcmd_box}.  
In both cases, we derived similar ages and age differences for the 
three fields.   
We note that this result differs from that reported in \citet{reb00} and
\citet{reb01}, where no age difference was found between optical samples
in the ONC and the Orion flanking fields.  
The earlier studies used larger fields and a different statistical method
that may have smeared out any age differences that might be present.
We note that none of the essential conclusions in earlier papers change
in any way if the Orion flanking fields are indeed older (or contain
a range of older ages) than the ONC. 

\section{CONCLUSIONS}

We present a catalog of Orion X-ray sources from two flanking fields.
The observations were taken with the ACIS-I on board the
Chandra X-ray Observatory.
The catalog, consisting of 417 sources, includes X-ray luminosity, 
optical and infrared photometry
and X-ray variability information. We found 91 variable sources,  
33 of which have a flare like light curve, and 11 of which
have a pattern of a steady increase or decrease. 
From the optical and infrared counterparts of the X-ray
sources, we have learned that most of the X-ray sources have 
colors consistent with CTTs that are younger than 
$10^{7}$ years. 
We argue that the data are consistent with an age difference among 
the X-ray selected samples of NGC 2264, Orion FF, and ONC, with 
NGC 2264 being the oldest, and ONC being the youngest.

This catalog of X-ray sources will
be used to study the relationship between rotational properties
and X-ray characteristics of Orion and NGC 2264 stars in paper III
\citep{reb04}.
We plan to discuss correlations of \lx/\lbol~ with
rotation rate (period and $v$sin$i$), disk indicators ($I_c-K$,
$H-K$, $U-V$, and $H\alpha$), and mass accretion rate as derived
from $U-V$ excess.  We will also compare the \lx/\lbol~ values
found here with those from other young clusters.
 
\acknowledgements
We thank the anonymous referee for her/his careful review of the
manuscript.
Financial support for this work was provided by NASA grant GO2-3011X.
This research has made extensive use of NASA's Astrophysics Data System
Abstract Service, the SIMBAD database, operated at CDS, Strasbourg, 
France, and the NASA/IPAC Infrared Science Archive, which is operated 
by the Jet Propulsion Laboratory, California Institute of Technology, 
under contract with the National Aeronautics and Space Administration.
The research described in this paper was partially carried out at
the Jet Propulsion Laboratory, California Institute of Technology,
under a contract with the National Aeronautics and Space Administration.

\clearpage



\clearpage
\begin{figure}
\epsscale{0.7}
\caption[dss.ps]{Image of Orion from the Palomar Digital Sky Survey
\citep{rei91}.
The image has a field of view of 60$\arcmin \times 90\arcmin$ and the 
boxes represent the fields of view of the Chandra ACIS-I observations. 
The North Orion Flanking Field (NOFF) is centered at 
RA(2000)=$5^h35^m19^s$, DEC(2000)=$-4\arcdeg48\arcmin15\arcsec$, and
the South Orion Flanking Field (SOFF) is centered at
RA(2000)=$5^h35^m6^s$, DEC(2000)=$-5\arcdeg40\arcmin48\arcsec$.
(figure available at http://spider.ipac.caltech.edu/staff/solange/ramirez07\_figs.ps)
\label{dss}}
\end{figure}

\begin{figure}
\epsscale{0.7}
\caption[orion_N.ps]{Image of the North Orion Flanking Field (NOFF), 
observed with ACIS-I at the Chandra Observatory. 
The image has a field of view of 17$\arcmin \times 17\arcmin$ 
and it is centered at RA(2000)=$5^h35^m19^s$, 
DEC(2000)=$-4\arcdeg48\arcmin15\arcsec$''.
This image contains only filtered events.
(figure available at http://spider.ipac.caltech.edu/staff/solange/ramirez07\_figs.ps)
\label{orion_N}}
\end{figure}

\begin{figure}
\epsscale{0.7}
\caption[orion_S.ps]{Image of the South Orion Flanking Field (SOFF), 
observed with ACIS-I at the Chandra Observatory. 
The image has a field of view of 17$\arcmin \times 17\arcmin$
and it is centered at RA(2000)=$5^h35^m6^s$, 
DEC(2000)=$-5\arcdeg40\arcmin48\arcsec$.
This image contains only filtered events.
(figure available at http://spider.ipac.caltech.edu/staff/solange/ramirez07\_figs.ps)
\label{orion_S}}
\end{figure}


\begin{figure}
\caption[spectra_N.ps]{Spectra of the 24 brightest sources in the 
NOFF sample.
The solid lines show the best two plasma temperature model.
(figure available at http://spider.ipac.caltech.edu/staff/solange/ramirez07\_figs.ps)
\label{spectra_N}}
\end{figure}

\begin{figure}
\epsscale{0.7}
\caption[spectra_S.ps]{Same as Fig.\ref{spectra_N}, but for the
20 brightest sources in the SOFF.
\label{spectra_S}}
(figure available at http://spider.ipac.caltech.edu/staff/solange/ramirez07\_figs.ps)
\end{figure}



\clearpage

\clearpage

\clearpage

\clearpage

\clearpage

\clearpage

\clearpage

\clearpage

%
%
\clearpage
\begin{deluxetable}{rcllcccccccrc}
\tabletypesize{\scriptsize}
\rotate
\tablenum{1}
\tablewidth{0pt}
\tablecaption{The Catalog of X-ray Sources\tablenotemark{a}. \label{tab_xray}}
\tablehead{
\colhead{X-ID} & 
\colhead{Name} &
\colhead{RA} &
\colhead{DEC} & 
\colhead{$\phi$} &
\colhead{$R_{ext}$} &
\colhead{$f_{PSF}$} &
\colhead{$t_{eff}$} &
\colhead{$C.~R.$} &
\colhead{Flux} &
\colhead{log(\lx)} &
\colhead{$P_{c}(\chi^{2})$} &
\colhead{Comments\tablenotemark{b}} \\ 
\colhead{} &
\colhead{} &
\colhead{(2000)} &
\colhead{(2000)} &
\colhead{(')} &
\colhead{(")} &
\colhead{} &
\colhead{(ks)} &
\colhead{(count/ks)} &
\colhead{(erg/cm$^{2}$/s)} &
\colhead{log(erg/s)} &
\colhead{\%} &
\colhead{} \\
\colhead{(1)} &
\colhead{(2)} &
\colhead{(3)} &
\colhead{(4)} &
\colhead{(5)} &
\colhead{(6)} &
\colhead{(7)} &
\colhead{(8)} &
\colhead{(9)} &
\colhead{(10)} &
\colhead{(11)} &
\colhead{(12)} &
\colhead{(13)}
}
\startdata 
\multicolumn{13}{c}{North Orion Flanking Field (NOFF)} \\ \hline
N001 & CXORRS J053442.6-044215 & 5 34 42.70 & -4 42 15.294 & 10.93 & 17.50 &  1.00 & 35.9 &  17.25 &  0.114E-12 & 30.48 &  93 &       \\
N002 & CXORRS J053444.5-044214 & 5 34 44.53 & -4 42 14.807 & 10.55 & 16.30 &  1.00 & 36.4 &  12.88 &  0.848E-13 & 30.35 &   0 & v,s   \\
N004 & CXORRS J053448.2-044740 & 5 34 48.24 & -4 47 40.865 &  7.77 &  9.50 &  0.98 & 40.9 &  36.05 &  0.237E-12 & 30.80 &   0 & v,s   \\
N005 & CXORRS J053448.5-044956 & 5 34 48.60 & -4 49 56.658 &  7.78 &  8.70 &  0.96 & 24.7 &   1.56 &  0.103E-13 & 29.43 &  99 &       \\
N007 & CXORRS J053451.0-044341 & 5 34 51.05 & -4 43 41.614 &  8.39 & 10.10 &  0.98 & 38.7 &  10.48 &  0.690E-13 & 30.26 &   0 & v,f   \\
N008 & CXORRS J053451.2-044757 & 5 34 51.28 & -4 47 57.405 &  7.00 &  9.40 &  0.99 & 41.6 &  55.40 &  0.365E-12 & 30.98 &  23 & v     \\
N010 & CXORRS J053452.4-044941 & 5 34 52.46 & -4 49 41.023 &  6.85 &  6.70 &  0.97 & 36.5 &   1.04 &  0.684E-14 & 29.26 & 100 &       \\
N011 & CXORRS J053453.0-044811 & 5 34 53.02 & -4 48 11.677 &  6.55 &  6.20 &  0.96 & 40.5 &   0.44 &  0.290E-14 & 28.88 & 100 &       \\
N012 & CXORRS J053453.8-044340 & 5 34 53.82 & -4 43 40.195 &  7.83 &  8.80 &  0.98 & 36.7 &   0.44 &  0.290E-14 & 28.88 & 100 &       \\
N015 & CXORRS J053454.3-045413 & 5 34 54.31 & -4 54 13.447 &  8.63 & 10.70 &  0.97 & 40.2 &   2.10 &  0.138E-13 & 29.56 &  97 &       \\
N016 & CXORRS J053454.4-044540 & 5 34 54.44 & -4 45 40.342 &  6.73 &  6.50 &  0.97 & 41.8 &   0.26 &  0.171E-14 & 28.66 & 100 &       \\
N017 & CXORRS J053454.5-045604 & 5 34 54.58 & -4 56  4.736 &  9.96 & 14.40 &  0.98 & 32.5 &   3.88 &  0.255E-13 & 29.83 &  99 &       \\
N018 & CXORRS J053455.1-044827 & 5 34 55.12 & -4 48 27.968 &  6.05 &  5.30 &  0.96 & 40.6 &  24.67 &  0.162E-12 & 30.63 &   2 & v     \\
N019 & CXORRS J053455.6-045611 & 5 34 55.67 & -4 56 11.579 &  9.89 & 14.20 &  0.98 & 39.3 &   3.78 &  0.249E-13 & 29.82 &  99 &       \\
N020 & CXORRS J053456.3-044548 & 5 34 56.31 & -4 45 48.316 &  6.25 &  5.70 &  0.97 & 37.7 &   1.14 &  0.750E-14 & 29.30 & 100 & x     \\
N021 & CXORRS J053456.3-044437 & 5 34 56.39 & -4 44 37.961 &  6.77 &  6.60 &  0.97 & 41.5 &   1.79 &  0.118E-13 & 29.49 & 100 & x     \\
N022 & CXORRS J053456.8-044605 & 5 34 56.81 & -4 46  5.044 &  6.01 &  5.30 &  0.97 & 39.8 &   8.12 &  0.534E-13 & 30.15 &  95 &       \\
N025 & CXORRS J053457.9-044913 & 5 34 57.90 & -4 49 13.139 &  5.43 &  4.40 &  0.95 & 19.6 &  14.61 &  0.961E-13 & 30.40 &  90 &       \\
N026 & CXORRS J053458.2-045052 & 5 34 58.22 & -4 50 52.031 &  5.86 &  5.00 &  0.96 & 40.1 &   0.27 &  0.178E-14 & 28.67 & 100 &       \\
N027 & CXORRS J053459.3-045011 & 5 34 59.35 & -4 50 11.837 &  5.36 &  4.30 &  0.95 & 42.7 &   2.40 &  0.158E-13 & 29.62 &  89 &       \\
N029 & CXORRS J053459.6-044756 & 5 34 59.63 & -4 47 56.143 &  4.92 &  3.70 &  0.95 & 43.4 &   0.68 &  0.447E-14 & 29.07 & 100 & x     \\
N030 & CXORRS J053459.7-045158 & 5 34 59.70 & -4 51 58.602 &  6.18 &  5.50 &  0.96 & 42.0 &   0.62 &  0.408E-14 & 29.03 &  98 &       \\
N031 & CXORRS J053459.8-045526 & 5 34 59.87 & -4 55 26.028 &  8.66 & 10.80 &  0.98 & 40.3 &   1.18 &  0.776E-14 & 29.31 & 100 &       \\
N034 & CXORRS J053500.7-044649 & 5 35  0.80 & -4 46 49.233 &  4.85 &  3.60 &  0.95 & 34.8 &   1.12 &  0.737E-14 & 29.29 &  99 &       \\
N035 & CXORRS J053500.9-044819 & 5 35  0.92 & -4 48 19.319 &  4.60 &  3.30 &  0.95 & 43.7 &   0.93 &  0.612E-14 & 29.21 &  98 &       \\
N037 & CXORRS J053501.9-044115 & 5 35  1.96 & -4 41 15.366 &  8.22 &  9.70 &  0.98 & 40.8 &   4.65 &  0.306E-13 & 29.91 &  13 & v     \\
N038 & CXORRS J053502.0-044731 & 5 35  2.09 & -4 47 31.843 &  4.36 &  3.00 &  0.95 & 34.9 &   1.14 &  0.750E-14 & 29.30 & 100 &       \\
N039 & CXORRS J053502.3-044755 & 5 35  2.33 & -4 47 55.508 &  4.21 &  2.90 &  0.95 & 32.3 &   2.64 &  0.174E-13 & 29.66 &  98 & x     \\
\enddata
\tablenotetext{a}{Table available electronically.}
\tablenotetext{b}{v: variable star; f: flare-like light curve; p: possible
flare in the light curve; s: steady increase or decrease in the light curve;
c: source confusion, see details in Sec.\ref{catalog};x: source detected only
in X-rays, see details in Sec.\ref{ccatalog}}
\end{deluxetable}

%
%
\clearpage
\begin{deluxetable}{cccccc}
\tabletypesize{\tiny}
\tablenum{2}
\tablewidth{0pt}
\tablecaption{Spectral Properties of Bright Sources. \label{tab_spectra}}
\tablehead{
\colhead{X-ID} &
\colhead{Optical ID} &
\colhead{$C.~R.$} &
\colhead{kT1} &
\colhead{kT2} &
\colhead{Flux\tablenotemark{a}} \\
\colhead{} &
\colhead{} &
\colhead{(counts/ks)} &
\colhead{(keV)} &
\colhead{(keV)} &
\colhead{($10^{-13}$erg/cm$^2$/s)}
}
\startdata
\multicolumn{6}{c}{North Orion Flanking Field (NOFF)} \\
 \hline
N093 &  Par 1817   & 209.4 & 0.77 & 3.40 & 14.20 \\
N256 &  Par 2257   &  65.6 & 0.70 & 2.87 &  4.03 \\
N218 &  Par 2140   &  60.6 & 0.56 & 1.96 &  3.61 \\
N087 &  Par 1798   &  56.0 & 1.05 & 3.62 &  3.78 \\
N008 &  Par 1621   &  55.4 & 0.40 & 2.63 &  3.60 \\
N097 &  Par 1834   &  52.5 & 0.00 & 3.58 &  3.35 \\
N126 &  Par 1950   &  43.5 & 0.65 & 2.22 &  2.65 \\
N044 &  Par 1701   &  35.4 & 0.00 & 2.74 &  2.54 \\
N004 &  Par 1598   &  36.0 & 0.00 & 2.11 &  2.21 \\
N137 &  Par 1967   &  30.3 & 0.31 & 1.39 &  1.88 \\
N165 &  Par 2043   &  41.4 & 0.30 & 1.31 &  2.54 \\
N124 &  Par 1935   &  27.8 & 0.00 & 1.23 &  1.65 \\
N190 &  Par 2081   &  29.5 & 0.85 & 4.40 &  1.77 \\
N222 &  Par 2145   &  27.0 & 0.61 & 2.40 &  1.54 \\
N018 &  Par 1651   &  24.7 & 0.84 & 3.88 &  1.60 \\
N195 &  R01 2133   &  21.1 & 0.40 & 3.25 &  1.25 \\
N080 &  Par 1778   &  20.3 & 0.00 & 4.02 &  1.16 \\
N208 &  Par 2109   &  17.0 & 0.24 & 1.07 &  0.91 \\
N081 & SMMV 1944   &  17.3 & 0.00 & 1.96 &  1.01 \\
N156 &  Par 2017   &  67.2 & 0.00 & 1.38 &  4.06 \\
N046 &  Par 1710   &  16.4 & 0.89 & 4.36 &  0.95 \\
N130 &  Par 1948   &  16.1 & 0.87 & 5.02 &  0.91 \\
N001 &  R01 1413   &  17.2 & 0.00 & 6.76 &  0.82 \\
N175 &  Par 2064   &  12.9 & 0.00 & 1.01 &  0.75 \\
 \hline
\multicolumn{6}{c}{South Orion Flanking Field (SOFF)} \\
 \hline
S249 &  Par 2069   & 200.9 & 0.85 & 3.71 & 13.30 \\
S168 &  Par 1828   & 100.8 & 1.00 & 4.80 &  7.21 \\
S238 &  Par 2048   & 103.8 & 0.66 & 3.34 &  7.25 \\
S020 &  Par 1553   &  52.0 & 0.70 & 2.43 &  3.40 \\
S198 &  Par 1929   &  37.5 & 0.66 & 2.23 &  2.41 \\
S040 &  Par 1756   &  34.7 & 0.77 & 2.70 &  2.16 \\
S183 &  Par 1874   &  31.6 & 1.00 & 4.45 &  2.13 \\
S152 &  Par 1787   &  27.3 & 1.33 & 5.82 &  1.85 \\
S058 &  Par 1643   &  26.5 & 0.64 & 2.02 &  1.34 \\
S172 &  Par 1846   &  21.2 & 0.64 & 2.65 &  1.26 \\
S202 &  Par 1942   &  19.5 & 0.78 & 2.66 &  1.12 \\
S174 &  Par 1848   &  19.5 & 0.86 & 3.70 &  1.14 \\
S022 &  Par 1564   &  17.8 & 0.64 & 1.63 &  1.12 \\
S047 &  Par 1613   &  16.0 & 0.69 & 1.76 &  0.86 \\
S016 &  Par 1535   &  15.4 & 0.64 & 1.63 &  0.85 \\
S039 &  Par 1616   &  15.6 & 0.64 & 1.47 &  0.85 \\
S176 &  Par 1876   &  14.4 & 0.81 & 3.36 &  0.86 \\
S079 &  R01 1603   &  13.1 & 0.32 & 3.58 &  0.64 \\
S025 &  Par 1571   &  13.4 & 1.04 & 1.88 &  0.84 \\
S104 &  CHS 7005   &  21.2 & 0.00 & 0.00 &  0.96 
\enddata
\tablenotetext{a}{Flux determined from models of mean plasma temperatures}
\end{deluxetable}

%
%
\clearpage
\begin{deluxetable}{cllrrrrrrr}
\tabletypesize{\small}
\tablenum{3}
\tablewidth{0pt}
\tablecaption{Optical/Infrared Sources with X-ray 
counterparts\tablenotemark{a}. \label{tab_optical}}
\tablehead{
\colhead{Name} &
\colhead{RA} &
\colhead{DEC} &
\colhead{X-ID} &
\colhead{U} &
\colhead{V} &
\colhead{${\rm I_c}$} &
\colhead{J} &
\colhead{H} &
\colhead{K} \\
\colhead{} &
\colhead{(2000)} &
\colhead{(2000)} &
\colhead{} &
\colhead{mag} &
\colhead{mag} &
\colhead{mag} &
\colhead{mag} &
\colhead{mag} &
\colhead{mag}
}
\startdata
  R01 1413              & 5 34 42.616  & -4 42 14.86  & N001 & \nodata &   18.79 &   15.52 &   13.07 &   10.99 &    9.94 \\
  Par 1567              & 5 34 44.512  & -4 42 13.71  & N002 &    0.04 &   15.38 &   13.53 &   12.36 &   11.49 &   10.89 \\
  Par 1598              & 5 34 48.155  & -4 47 39.98  & N004 &    0.11 &   16.63 &   14.08 &   12.05 &   10.73 &    9.97 \\
  CHS 5807              & 5 34 48.548  & -4 49 56.72  & N005 & \nodata & \nodata & \nodata & \nodata &   13.97 &   12.07 \\
  Par 1620              & 5 34 50.988  & -4 43 41.35  & N007 & \nodata &   16.02 &   13.60 &   12.14 &   11.09 &   10.43 \\
  Par 1621              & 5 34 51.204  & -4 47 56.91  & N008 & \nodata &   15.85 &   13.45 &   11.88 &   10.66 &    9.96 \\
  R01 1536              & 5 34 52.410  & -4 49 40.29  & N010 & \nodata & \nodata &   19.34 &   16.40 &   13.76 &   12.45 \\
  Par 1639              & 5 34 52.963  & -4 48 10.56  & N011 & \nodata & \nodata &   17.10 &   15.20 & \nodata &   12.23 \\
  R01 1563              & 5 34 53.670  & -4 43 40.69  & N012 & \nodata & \nodata &   18.12 &   15.82 &   14.57 &   13.95 \\
2MASS J05345431-0454129 &  5 34 54.313 & -4 54 13.447 & N015 & \nodata & \nodata & \nodata & \nodata &   14.81 &   12.42 \\
  R01 1570              & 5 34 54.391  & -4 45 39.22  & N016 & \nodata & \nodata &   19.70 &   16.11 &   13.83 &   13.11 \\
  R01 1572              & 5 34 54.501  & -4 56  4.96  & N017 & \nodata & \nodata &   18.05 &   14.65 &   12.11 &   10.93 \\
  Par 1651              & 5 34 55.052  & -4 48 27.54  & N018 &    0.47 &   16.27 &   14.12 &   12.44 &   11.13 &   10.41 \\
  R01 1586              & 5 34 55.574  & -4 56 11.23  & N019 & \nodata & \nodata &   19.11 &   15.25 &   12.47 &   11.00 \\
  CHS 6568              & 5 34 56.817  & -4 46  4.78  & N022 & \nodata & \nodata & \nodata & \nodata & \nodata &   11.46 \\
  R01 1610              & 5 34 57.827  & -4 49 12.54  & N025 & \nodata & \nodata &   16.99 &   14.43 &   12.73 &   11.91 \\
  R01 1620              & 5 34 58.184  & -4 50 51.45  & N026 & \nodata & \nodata &   18.61 &   15.72 &   13.30 &   12.38 \\
  Par 1672              & 5 34 59.269  & -4 50 11.41  & N027 & \nodata & \nodata &   15.85 &   13.38 &   11.77 &   10.93 \\
  R01 1635              & 5 34 59.636  & -4 51 57.50  & N030 & \nodata & \nodata &   21.60 &   16.58 &   13.94 &   12.94 \\
2MASS J05345988-0455272 &  5 34 59.874 & -4 55 26.028 & N031 & \nodata & \nodata & \nodata & \nodata &   17.05 &   14.02 \\
  R01 1647              & 5 35  0.740  & -4 46 49.07  & N034 & \nodata & \nodata &   16.52 &   13.99 &   12.44 &   11.69 \\
  R01 1651              & 5 35  0.846  & -4 48 18.71  & N035 & \nodata & \nodata &   17.23 &   14.22 &   12.38 &   11.58 \\
  R01 1659              & 5 35  1.890  & -4 41 14.54  & N037 &    1.35 & \nodata &   18.86 &   15.55 &   13.22 &   12.14 \\
  R01 1660              & 5 35  2.000  & -4 47 30.73  & N038 & \nodata & \nodata &   21.16 &   15.93 &   13.49 &   12.86 \\
  Par 1692              & 5 35  2.303  & -4 49 15.92  & N040 &    0.24 & \nodata &   16.74 &   14.28 &   12.70 &   11.88 \\
  R01 1663              & 5 35  2.574  & -4 49 29.12  & N042 & \nodata &   17.68 &   14.87 &   12.60 &   11.09 &   10.22 \\
  Par 1700              & 5 35  2.891  & -4 48 32.54  & N043 & \nodata &   16.57 &   16.52 &   14.61 &   13.21 &   12.51 \\
  Par 1701              & 5 35  3.229  & -4 49 20.29  & N044 & \nodata &   14.31 &   13.01 &   11.32 &    9.96 &    9.08 \\
  R01 1677              & 5 35  3.252  & -4 56 42.63  & N045 & \nodata &   18.24 &   15.60 &   13.58 &   12.33 &   11.57 \\
  Par 1710              & 5 35  3.622  & -4 50 52.77  & N046 & \nodata &   16.47 &   15.11 &   13.19 &   11.47 &   10.36 \\
2MASS J05350376-0447516 &  5 35  3.748 & -4 47 51.576 & N047 & \nodata & \nodata & \nodata & \nodata &   14.24 &   13.47 \\
  Par 1709              & 5 35  4.292  & -4 46 42.20  & N048 & \nodata & \nodata & \nodata & \nodata &    9.79 &    9.26 \\
2MASS J05350469-0452418 &  5 35  4.668 & -4 52 41.560 & N049 & \nodata & \nodata & \nodata & \nodata &   14.35 &   12.67 \\
\enddata
\tablenotetext{a}{Table available electronically.}
\end{deluxetable}

%
%
\clearpage
\begin{deluxetable}{rlllll}
\tablenum{4}
\tabletypesize{\scriptsize}
\tablewidth{0pt}
\tablecaption{Cross identification of sources with optical/infrared 
counterparts\tablenotemark{a}. \label{tab_xids}}
\tablehead{
\colhead{X-ID} &
\colhead{Par name\tablenotemark{b}} &
\colhead{R01 name\tablenotemark{c}} &
\colhead{CHS name\tablenotemark{d}} &
\colhead{2MASS name\tablenotemark{e}} &
\colhead{Other names\tablenotemark{f}} 
}
\startdata
N001 & \nodata   & R01 1413  & CHS 5282   & 2MASS J05344268-0442148 &                                                    SMMV 1089 \\
N002 & Par 1567  & R01 1429  & CHS 5438   & 2MASS J05344454-0442146 &                                  Tian 125,ROSAT 42,SMMV 1126 \\
N004 & Par 1598  & R01 1483  & \nodata    & 2MASS J05344823-0447401 &                                                     ROSAT 52 \\
N005 & \nodata   & \nodata   & CHS 5807   & 2MASS J05344854-0449568 &                                                   \nodata    \\
N007 & Par 1620  & R01 1521  & \nodata    & 2MASS J05345106-0443414 &                                                   \nodata    \\
N008 & Par 1621  & R01 1522  & CHS 6064   & 2MASS J05345128-0447570 &                                         Einstein 56,ROSAT 59 \\
N010 & \nodata   & R01 1536  & \nodata    & 2MASS J05345249-0449404 &                                                   \nodata    \\
N011 & Par 1639  & R01 1542  & \nodata    & 2MASS J05345302-0448104 &                                                   \nodata    \\
N012 & \nodata   & R01 1563  & \nodata    & 2MASS J05345374-0443407 &                                                   \nodata    \\
N015 & \nodata   & \nodata   & \nodata    & 2MASS J05345431-0454129 &                                                    \nodata   \\
N016 & \nodata   & R01 1570  & \nodata    & 2MASS J05345446-0445393 &                                                   \nodata    \\
N017 & \nodata   & R01 1572  & \nodata    & 2MASS J05345458-0456053 &                                                   \nodata    \\
N018 & Par 1651  & R01 1578  & CHS 6412   & 2MASS J05345513-0448277 &                                                     ROSAT 72 \\
N019 & \nodata   & R01 1586  & CHS 6469   & 2MASS J05345566-0456117 &                                                   \nodata    \\
N022 & \nodata   & \nodata   & CHS 6568   & 2MASS J05345682-0446047 &                                                   \nodata    \\
N025 & \nodata   & R01 1610  & \nodata    & 2MASS J05345791-0449127 &                                                    SMMV 1452 \\
N026 & \nodata   & R01 1620  & \nodata    & 2MASS J05345826-0450517 &                                                   \nodata    \\
N027 & Par 1672  & R01 1633  & \nodata    & 2MASS J05345935-0450117 &                                                   \nodata    \\
N030 & \nodata   & R01 1635  & \nodata    & 2MASS J05345971-0451578 &                                                   \nodata    \\
N031 & \nodata   & \nodata   & \nodata    & 2MASS J05345988-0455272 &                                                    \nodata   \\
N034 & \nodata   & R01 1647  & \nodata    & 2MASS J05350082-0446491 &                                                   \nodata    \\
N035 & \nodata   & R01 1651  & \nodata    & 2MASS J05350093-0448188 &                                                   \nodata    \\
N037 & \nodata   & R01 1659  & CHS 7084   & 2MASS J05350196-0441145 &                                                   \nodata    \\
N038 & \nodata   & R01 1660  & \nodata    & 2MASS J05350208-0447308 &                                                   \nodata    \\
N040 & Par 1692  & R01 1662  & \nodata    & 2MASS J05350238-0449161 &                                                   \nodata    \\
N042 & \nodata   & R01 1663  & \nodata    & 2MASS J05350265-0449293 &                                                   \nodata    \\
N043 & Par 1700  & R01 1668  & CHS 7175   & 2MASS J05350298-0448326 &                                                   \nodata    \\
N044 & Par 1701  & R01 1676  & CHS 7211   & 2MASS J05350326-0449209 &                                                     Tian 158 \\
N045 & \nodata   & R01 1677  & \nodata    & 2MASS J05350333-0456430 &                                                   \nodata    \\
N046 & Par 1710  & R01 1678  & CHS 7251   & 2MASS J05350370-0450530 &                                                   \nodata    \\
N047 & \nodata   & \nodata   & \nodata    & 2MASS J05350376-0447516 &                                                    \nodata   \\
N048 & Par 1709  & \nodata   & \nodata    & 2MASS J05350421-0446435 &                                                     Tian 161 \\
N049 & \nodata   & \nodata   & \nodata    & 2MASS J05350469-0452418 &                                                    \nodata   \\
N050 & \nodata   & \nodata   & \nodata    & 2MASS J05350481-0447089 &                                                    \nodata   \\
\enddata
\tablenotetext{a}{Table available electronically.}
\tablenotetext{b}{\citet{par54}.}
\tablenotetext{c}{\citet{reb01}.}
\tablenotetext{d}{\citet{car01}.}
\tablenotetext{e}{2MASS Catalog.}
\tablenotetext{f}{SMMV (\citet{sta99}); Tian (\citet{tia96}); ROSAT (\citet{gag94}); Einstein (\citet{gag95}); 
HBC (\citet{her88}); JW (\citet{jon88}); H97 (\citet{hil97}); Feigelson (\citet{fei02}); HBJM (\citet{her01}).}
\end{deluxetable}

%
%
\clearpage
\begin{deluxetable}{lllllccccc}
\tabletypesize{\scriptsize}
\rotate
\tablenum{5}
\tablewidth{0pt}
\tablecaption{Upper limits for some optical/infrared stars in the
Chandra field. \label{tab_upp_lim}}
\tablehead{
\colhead{Par Name} &
\colhead{R Name} &
\colhead{CHS Name} &
\colhead{2MASS Name} &
\colhead{Other Names} &
\colhead{Counts} &
\colhead{$C.~R.$} &
\colhead{Flux} &
\colhead{log(\lx)} \\
\colhead{} &
\colhead{} &
\colhead{} &
\colhead{} &
\colhead{} &
\colhead{(counts)} &
\colhead{(counts/ks)} &
\colhead{($10^{-13}$erg/cm$^2$/s)} &
\colhead{log(erg/s)}
}
\startdata
\nodata   & \nodata   & CHS 10870  & \nodata                 &                                                   \nodata    & $<$ 6.30 &
$<$ 0.3447 & $<$0.0227 & $<$ 28.78 \\
Par 2083  & \nodata   & \nodata    & \nodata                 &                                                     Tian 250 & $<$10.70 &
$<$ 0.4994 & $<$0.0329 & $<$ 28.94 \\
Par 2131  & \nodata   & \nodata    & 2MASS J05353948-0451216 &                                                     Tian 258 & $<$ 9.80 &
$<$ 0.2455 & $<$0.0161 & $<$ 28.63 \\
Par 1654  & R01 1599  & \nodata    & 2MASS J05345622-0445574 &                                                     Tian 150 & $<$ 8.00 &
$<$ 0.2016 & $<$0.0133 & $<$ 28.54 \\
Par 1708  & \nodata   & \nodata    & 2MASS J05350478-0443546 &                                                     Tian 162 & $<$ 8.40 &
$<$ 0.2328 & $<$0.0153 & $<$ 28.61 \\
\nodata   & \nodata   & CHS 8034   & 2MASS J05351065-0442075 &                                                   \nodata    & $<$ 6.30 &
$<$ 0.1464 & $<$0.0096 & $<$ 28.41 \\
\nodata   & \nodata   & CHS 8467   & 2MASS J05351406-0453112 &                                                   \nodata    & $<$ 5.10 &
$<$ 0.1671 & $<$0.0110 & $<$ 28.46 \\
\nodata   & R01 1937  & CHS 9321   & 2MASS J05351974-0448180 &                                                   \nodata    & $<$ 3.00 &
$<$ 0.1881 & $<$0.0124 & $<$ 28.51 \\
\nodata   & R01 1720  & \nodata    & 2MASS J05350693-0449097 &                                                   \nodata    & $<$13.90 &
$<$ 0.8425 & $<$0.0554 & $<$ 29.17 \\
\nodata   & R01 1806  & \nodata    & 2MASS J05351282-0539077 &                                                       JW 398 & $<$ 3.00 &
$<$ 0.2237 & $<$0.0150 & $<$ 28.60 \\
\nodata   & \nodata   & CHS 9056   & 2MASS J05351795-0535157 &                                             JW 579,SMMV 2470 & $<$ 6.70 &
$<$ 0.1530 & $<$0.0103 & $<$ 28.43 \\
\nodata   & \nodata   & \nodata    & 2MASS J05351898-0537234 &                                                   HBJM 10548 & $<$ 3.80 &
$<$ 0.1427 & $<$0.0096 & $<$ 28.40 \\
\nodata   & \nodata   & \nodata    & \nodata                 &                                                       JW 618 & $<$ 5.60 &
$<$ 0.1889 & $<$0.0127 & $<$ 28.53 \\
Par 1810  & R01 1795  & \nodata    & 2MASS J05351235-0536403 &                                                       JW 384 & $<$10.10 &
$<$ 0.4175 & $<$0.0280 & $<$ 28.87 \\
Par 1898  & \nodata   & \nodata    & 2MASS J05351596-0539147 &                                              JW 514,Tian 196 & $<$ 3.00 &
$<$ 0.1990 & $<$0.0134 & $<$ 28.55 \\
\nodata   & R01 1949  & \nodata    & 2MASS J05352074-0537536 &                                                       JW 667 & $<$ 8.50 &
$<$ 0.3344 & $<$0.0225 & $<$ 28.77 \\
\nodata   & R01 1451  & \nodata    & 2MASS J05344609-0537312 &                                                        JW 79 & $<$12.10 &
$<$ 0.3498 & $<$0.0235 & $<$ 28.79 \\
\enddata
\end{deluxetable}

%
%
\clearpage
\begin{deluxetable}{cccccccccc}
\rotate
\tablenum{6}
\tablewidth{0pt}
\tablecaption{Median and Quartile Values for NGC 2264, 
Orion Flanking Fields, and Orion Nebula Cluster \label{tab_boxes}}
\tablehead{
\colhead{} & \multicolumn{3}{c}{NGC 2264} &
\multicolumn{3}{c}{Orion Flanking Fields} & 
\multicolumn{3}{c}{Orion Nebula Cluster} \\
\colhead{Color Range} & 
\colhead{Upper } &
\colhead{Median} &
\colhead{Lower} &
\colhead{Upper} &
\colhead{Median} &
\colhead{Lower} &
\colhead{Upper} &
\colhead{Median} &
\colhead{Lower} \\
\colhead{} & 
\colhead{Quartile} &
\colhead{} &
\colhead{Quartile} &
\colhead{Quartile} &
\colhead{} &
\colhead{Quartile} &
\colhead{Quartile} &
\colhead{} &
\colhead{Quartile}
}
\startdata
1.0--1.5 & 3.81 & 4.21 & 4.44 & 3.18 & 3.71 & 4.22 & 2.66 & 3.43 & 3.90 \\
1.5--2.0 & 4.30 & 4.88 & 5.25 & 4.15 & 4.53 & 4.88 & 3.63 & 4.30 & 4.69 \\
2.0--2.5 & 5.10 & 5.58 & 5.98 & 4.81 & 5.30 & 5.72 & 4.57 & 4.94 & 5.45 \\
\enddata
\end{deluxetable}

\end{document}